# Two binary stars gravitational waves – homotopy perturbation method

M Zare *, O jalili, M Delshadmanesh
Department of Physics
Islamic Azad
University
(Nour Branch), IRAN
PIN code:4817935861

**Abstract :** Homotopy perturbation is one of the newest methods for numerical analysis of deferential equations. We have used for solving wave equation around a black hole. Our conclusions have this method far reaching consequences for comparison of theoritical physics and experimental physics.

* E-mail : < a1148730@unet.univie.ac.at >

*1. Introduction*

The stars in the sky are usually binary and our sun is an exception among them. The binary stars rotate around each other, otherwise they collapse onto each other. Since the mass of the stars is very high, the binary stars are the most important source of gravitational waves(GW). Through those two binary neutron stars and black holes binary produce stronger gravitational waves. Thus, these are studied as the main candidate in sky observations for GW [1]. The gravitational waves emission reduces system's energy and causes the reduction of the distance between the two stars. As a consequence, the stars will rotate with more speed around each other. The higher speeds, the higher GW's frequency. (LIGO) and (VIRGO) detectors works in rang 10 HZ – 100 HZ [2]. Indeed this frequency range is the final step of binary stars before they collapse.

In this step, speed of stars is very high ,so we should use modified Newton's equations or general relativity as well. The GW of binary stars can give us useful information such as direction of binary system, position in the sky, mass, spin and the distance between them.

The most important equation of binary stars that we have here on the earth is the dependence of system's energy on their radiation frequency. It can be shown that the energy decrease rate is as follows [1].

$$E = E_{QF}[1 + O(v^2) + O(v^3) + O(v^4) + ...]  \quad (1)$$

In above equation $E_{QF} = (32/5)(\mu/M)^2 v^{10}$ is the radiation's main term, $\mu$ and $M$ are respectively the reduced mass and the mass of the whole system and also $v = (\pi M f)^{1/3}$.

The velocity in this equation is assumed to be the velocity that produces the low frequency range of gravitational spectrum( 10 Hz).

Each term on this expansion is in fact related to a special effect. The term $O(v^2)$ is related to Newtonian high effects [3]. The term $O(v^3)$ is related to two effects: first wave tail effect [4] and the second, spin-orbital interaction effect [5]. The term $O(v^4)$ is related to spin-spin effects [5].

Perturbation analysis can be used when the mass of one of the two stars relative to whole mass be very small (rotation of one star around a black hole). In section 2, we will discuss gravitational field's perturbation of static or rotating black hole and in section 3, we introduce homotopy perturbation method and then in section 4, we will apply this method to the problems of the section 2 and then compare our numerical result with other numerical analysis.

## 2. Perturbation of Black Hole 's Gravitational Field

The gravitational field of a Schwarzschild black hole is defined by the following metric:

$$ds^2 = -(1 - \frac{2M}{r})dt^2 + (1 - \frac{2M}{r})^{-1}dr^2 + r^2(d\theta^2 + \sin^2\theta d\varphi^2) \quad (2)$$

The wave equation in the Schwarzschild background is:

$$\Box \Phi \equiv (-g)^{-1/2} \partial_\mu [(-g)^{-1/2} g^{\mu\nu} \partial_\nu \Phi] = 0 \tag{3}$$

where $g$ is the determinant of the metric tensor $g_{\mu\nu}$. GW is a spin two field, thus we must consider s=2 field. Now we consider for simplicity, massless scalar field first. Since the metric of the Schwarzschild black hole has spherical symmetry we guess, the solution has the following form:

$$\Phi_{lm} = \frac{u_l(r,t)}{r} Y_{lm}(\theta, \Phi) \tag{4}$$

Where $Y_{lm}(\theta, \Phi)$ are the spherical harmonics. By substituting Eq.(4) into Eq.(3) we find that the function $u_l(r,t)$ satisfies the following equation

$$[\frac{\partial^2}{\partial r_*^2} - \frac{\partial^2}{\partial t^2} - v_l(r)] u_l(r,t) = 0 \tag{5}$$

Where $v_l(r)$ is defined by

$$v_l(r) = (1 - \frac{2M}{r}) \left[ \frac{l(l+1)}{r^2} + \frac{2M}{r^3} \right] \tag{6}$$

And $r_*$ is defined by

$$\frac{d}{dr_*} = (1 - \frac{2M}{r}) \frac{d}{dr} \tag{7}$$

Or equivalently as

$$r_* = r + 2M \log(\frac{r}{2M} - 1) + \text{constant} \tag{8}$$

The $r_*$ coordinate is called tortoise coordinate because the diagram of $v_l(r)$ in $r_*$ has tortoise shape [6]. Fig.1(a) shows the effective potential in term of $r$ and Fig.1(b) shows the effective potential in term of $r_*$.

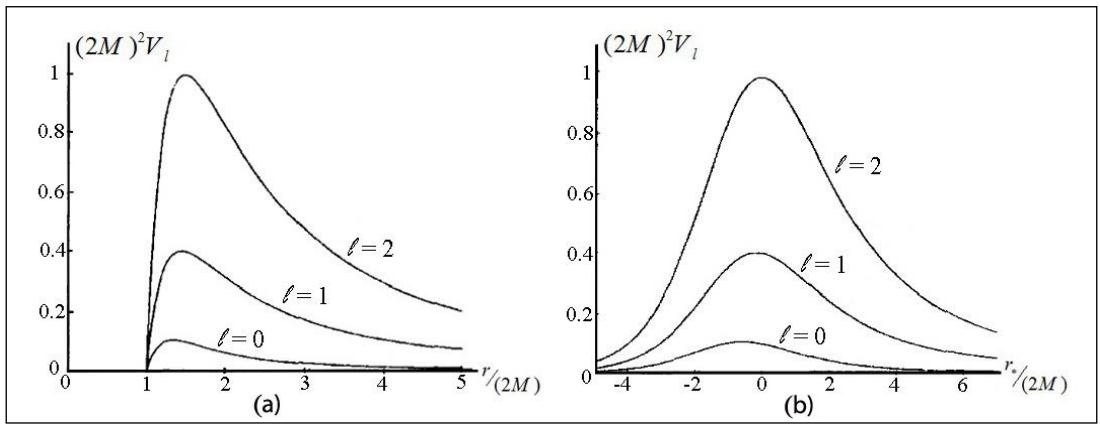

Figure 1: The effective potential $V_\ell$ for $\ell = 0, 1, 2$ as the function of $r$ Fig.1(a) and $r_*$ Fig.1(b) [7].

In those figure $r_* \to +\infty$ is similar to $r \to +\infty$ in Schwarzschild coordinate, whereas $r_* \to -\infty$ is similar to $r \to 2M$ namely the event horizon of the black hole. The potential vanishes for both of far distance and the event horizon. Then for these regions, the field satisfies the flat space wave equation. This equation in the time domain has the form

$$[\frac{d^2}{dr_*^2} + w^2 - v_l(r)]u_l(r,w) = 0 \tag{9}$$

Since this equation has the same form as that obtained by Regge and Wheeler for the first time for gravitational perturbations in 1957, is called Regge-Wheeler equation.

Now we discuss other spins. Prise showed that the waves equation with spin S is the same as the Regge-Wheeler equation with the replaced potential [8,9]:

$$v_l^{(s)}(r) = (1 - \frac{2M}{r})\left[\frac{l(l+1)}{r^2} - \frac{2M(1-S^2)}{r^3}\right] \tag{10}$$

For our purpose in this paper, we have, $S = 2$. Then the effective potential has the following form

$$v_l^{(s=2)}(r) = (1 - \frac{2M}{r})\left[\frac{l(l+1)}{r^2} - \frac{6M}{r^3}\right] \tag{11}$$

Exact solution of wave equation with S=2 shows that there is another solution with the following potential:

$$v_l(r) = 2(1 - \frac{2M}{r}) \frac{n^2(n+1)r^3 + 3n^2Mr^2 + 9nM^2r + 9M^3}{r^3(nr+3M)^2} \qquad (12)$$

This potential for the first times was proposed by Zerilli in 1970 [10]. Chanrasekhar called the first one axial and the second one polar [11].The Chanrasekhar;s terminology is more common.

If a Gaussian wave pocket from infinity strike to the black hole (namely to the potential of fig.1(a) or fig.1 (b) then the reflected wave behave like fig.2 .

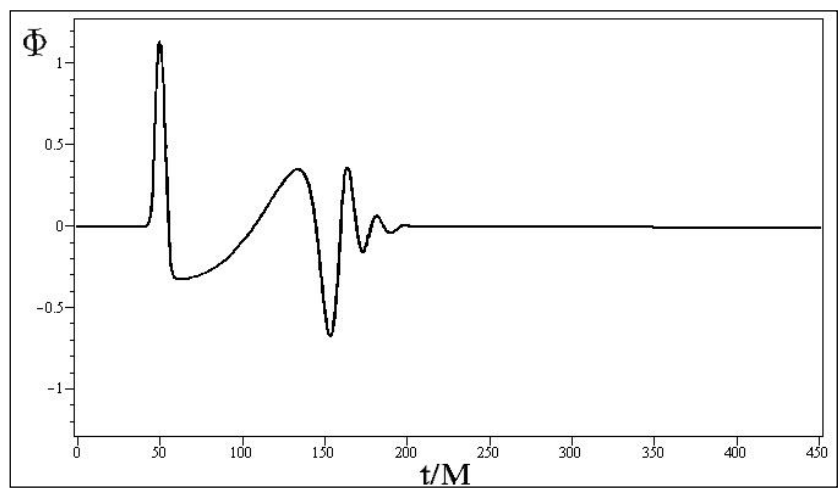

Figure 2: The response of a Schwarzschild black hole to a Gaussian wave packet [7].

This behavior is a reflection of black hole's nature. If a particle falls in the black hole or if a star collapses in it selves to form a black hole, we would be encountered with the behavior like a scattering problem. Thus in this paper, we choose the scattering model and we analyze Eqs.(5) and (9) by using homotopy perturbation technique for Gaussian wave scattering [12,13].

## 4. Solution of the Regge-Wheeler Equation using homotopy perturbation method

Homotopy perturbation method (HPM) was first proposed by He [12] and during the years it was applied to various engineering and scientific problems [21-24]. In this section we first extent the meaning of the two homotopic function to the two homotopic differential equations and then construct the whole theory and at the end we apply HPM to our problem.

If two functions $f$ and $g$ be homotope then curves between them can be chosen by different methods. In general, all different paths can be obtained by selecting a proper class of $g_i$ functions, from the following expansion.

$$G(x,p) = g_0(x) + g_1(x)p + g_2(x)p^2 + \ldots \qquad (13)$$

Where $g_0(x)$ is the same as $f(x)$ and $g(x,1) = g(x)$. If two functions $f$ and $g$ be homotop then $L_1(f)$ and $L_2(g)$ will be homotop, where $L_1$ and $L_2$ are two arbitrary operator. Therefore, we can use the homotopy method to solve differential Eq.(9). Suppose $f$ be a solution of following differential equation.

$$L_1(f) = 0 \qquad (14)$$

And $g$ be a solution of following differential equation.

$$L_2(g) = 0 \qquad (15)$$

Suppose that $f$ and $g$ are homotope and also $L_1(f)$ and $L_2(g)$ be homotope, so one can construct following homotopy between them:

$$H(x,p) = (1-p)L_1(f) + pL_2(g) \qquad (16)$$

We see $H(x,0) = L_1(f)$ and $H(x,1) = L_2(g)$. Then in these two limits, the H must vanish so:

$$H(x,0) = 0 \quad \& \quad H(x,1) = 0 \qquad (17)$$

Without losing generality, we can equal intermediate functions to zero, therefore

$$H(x,p) = (1-p)L_1(f) + pL_2(g) = 0 \tag{18}$$

For any $x \in \Omega$ and $p \in [0,1]$ where $\Omega$ is domain of $x$. We have to notice that if we choose the special homotopy Eq.(16) for state the homotopy between $L_1(f)$ and $L_2(g)$ Then we can use the general homotopy Eq.(13) to state the homotopy between $f$ and $g$. Therefore, Eq.(18) will be a series in $p$. We will get unknown function [Eq.(13)] with vanishing coefficients of this expansion. We usually know The solution of equation $L_1(f) = 0$ and want to obtain the solution of equation $L_2(g) = 0$. In most problems $L_1$ is a part of $L_2$. For example, $L_1$ is a linear part of $L_2$ operator [14-20].

For solving of Eq.(5), by using homotopy perturbation method, we suppose the $L_1$ operator to be of following form.

$$L_1 = \frac{\partial^2}{\partial r_*^2} - \frac{\partial^2}{\partial t^2} \tag{19}$$

It is clear that the we know the solutions of $L_1(f) = 0$. This is the wave equation that one of the familiar solutions of it is the Gaussian wave.

$$f(x,t) = e^{-(r_0 + x + t)^2} \tag{20}$$

Then we consider $V_1(r)$ potential as a perturbation for wave equation $L_1(f) = 0$, this means the $L_2$ operator is as follows

$$L_2 = \frac{\partial^2}{\partial r_*^2} - \frac{\partial^2}{\partial t^2} - V_1(r) \tag{21}$$

Therefore, if $u(r,t)$ be a solution of the equation $L_2(u) = 0$, we can write for it the following homotopy expansion similar to Eq.(14).

$$u(r,t) = u_0(r,t) + u_1(r,t)p + u_2(r,t)p^2 + u_3(r,t)p^3 + \ldots \tag{22}$$

Where $u_0(r,t) = e^{-(r_0 + x + t)^2}$.

Here, we also supposed that a Gaussian wave come to black hole from a far primary position $r_0$. By substituting expansion Eq.(23) into the following homotopy

$$H = \left[(1-p)(\frac{\partial^2}{\partial r_*^2} - \frac{\partial^2}{\partial t^2}) + p(\frac{\partial^2}{\partial r_*^2} - \frac{\partial^2}{\partial t^2} - V_1(r))\right]u = 0 \quad (23)$$

We get a power series in term of P which coefficients of the first four terms of it is as follows

$$p^0 : (\frac{\partial^2}{\partial r_*^2} - \frac{\partial^2}{\partial t^2})u_0 = 0 \quad (24)$$

$$p^1 : (\frac{\partial^2}{\partial r_*^2} - \frac{\partial^2}{\partial t^2})u_1 - (1 - \frac{2M}{r})(\frac{l(l+1)}{r^2} + \frac{6M}{r^3})u_0 = 0 \quad (25)$$

$$p^2 : (\frac{\partial^2}{\partial r_*^2} - \frac{\partial^2}{\partial t^2})u_2 - (1 - \frac{2M}{r})(\frac{l(l+1)}{r^2} + \frac{6M}{r^3})u_1 = 0 \quad (26)$$

$$p^3 : (\frac{\partial^2}{\partial r_*^2} - \frac{\partial^2}{\partial t^2})u_3 - (1 - \frac{2M}{r})(\frac{l(l+1)}{r^2} + \frac{6M}{r^3})u_2 = 0 \quad (27)$$

Using the arrangement in these equations, we can easily write upper coefficients. As one can see from the first term, Eq.(25) is nothing just nonperturbed equation, which we used the Gaussian solution of it. Substituting solution $u_0$ into Eq.(26), we can simply solve it. Notice that Eq.(26) is the same as non-homogen wave equation, which we have known its solution. Having $u_0$, $u_1$ can be obtained by solving the Eq(27) and so on.

And at the end the final solution will be obtained by adding perturbation terms for $p = 1$.

$$u(r,t) = u_0 + u_1 + u_3 + u_4 \quad (28)$$

Notice that for simplicity we used $r_*$ instead of $r$. We have used Maple13 for numeric computations, the software results depicted as a dashed curve in Figure (3) that have very good match with other numerical computations (Fig.3).

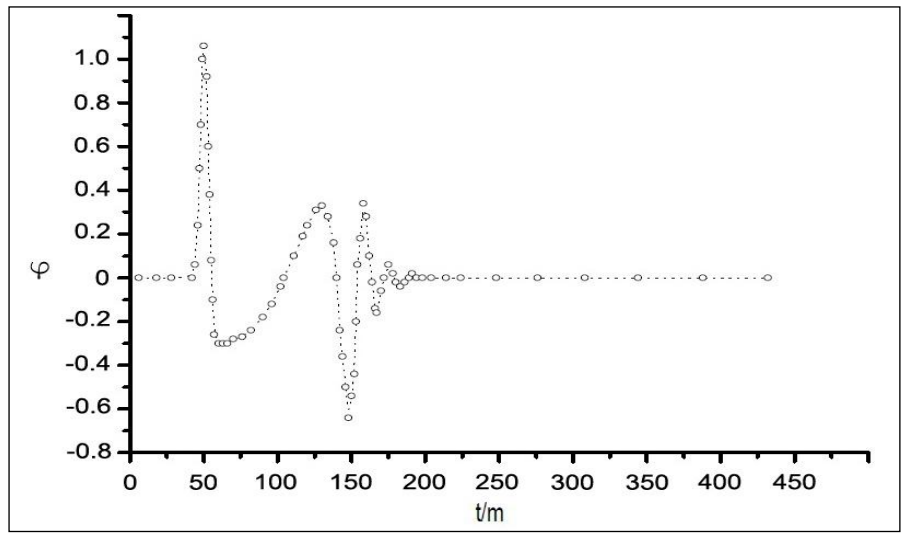

Figure 3: Response of a Schwarzschild black hole to a Gaussian wavepacket , using homotopy perturbation method .

## 5. Conclusions

The homotopy perturbation method as a useful method for solving differential equation is used in the problem of gravitational radiation of binary stars. We have found this method is very useful to explain the bell frequency of black hole. This frequency is very interested for black hole research. Thus by fine analyzing of the tail of figure (2) we can evaluate the validity of homotopy method.